\documentclass[a4paper,10pt]{article}
\usepackage[pdftex]{graphicx}
\pdfoutput=1
\usepackage{amsmath}
\usepackage{accents}
\usepackage[rflt]{floatflt}
\usepackage{color,calc}
\usepackage{wrapfig}

\newcommand{\commentout}[1]{}

\newcommand{\onlystats}[1]{}
\newcommand{\toomuchstats}[1]{#1}
\newcommand{\fabulous}[1]{#1}

\newcommand{\mysubsection}[1]{\section{#1}}
\onlystats{
\renewcommand{\mysubsection}[1]{\subsection{#1}}
\usepackage{shadbox}
}

\newcommand{\showguide}[1]{{\em #1}}
\newcommand{\shortrefs}[1]{#1}

\newcommand{\pastt}{\kappa}
\newcommand{\futt}{\tau}
\renewcommand{\showguide}[1]{}
\renewcommand{\shortrefs}[1]{}

\definecolor{shade}{gray}{0.8}

\usepackage{helvet}

\title{The Two-Way Likelihood Ratio (G) Test \\ {\small and Comparison to Two-Way $\chi^2$ Test}}
\onlystats{
\title{Creativity: A Mathematical Definition \\ {\small Based on The Two-Way Likelihood Ratio (G) Test}}
}
\author{Jesse Hoey}
\begin{document}
\maketitle
\toomuchstats{
\mysubsection{One-Way Likelihood Ratio or $\chi^2$ test}

Suppose we have a set of data $\mathbf{x}$ and two hypotheses $H_R$ and $H_S$. We wish to
know which hypothesis explains the data better.  To do this, we compute the likelihood ratio
\[ \log\left(\frac{P(\mathbf{x}|H_R)}{P(\mathbf{x}|H_S)}\right)\]

Assuming the data are i.i.d given each hypothesis, we have $P(\mathbf{x}|H_J)=\prod_i P(x_i|H_J)$, where $J\in{R,S}$,
and thus the likelihood ratio is 
\begin{equation}
L= \sum_i \log\left(\frac{P(x_i|H_R)}{P(x_i|H_S)}\right)
\label{lrat1}
\end{equation}

The Bayesian formulation of the problem could be approached by parameterising $H_R$ and $H_S$ with some unknown parameters, $\theta_R$ and $\theta_S$, respectively.  The posterior distribution over these parameters is then given by integrating the likelihoods over all possible values
\begin{align}
L=\frac{\log\left(\int P(\theta_R|H_R)P(\mathbf{x}|\theta_R,H_R)d\theta_R\right)}{\log\left(\int P(\theta_S|H_S)P(\mathbf{x}|\theta_S,H_s)d\theta_S\right)}
\label{likeratbayes1}
\end{align}
These integrations can sometimes be performed analytically, or using some numerical integration techniques.  However, we will focus instead on a simple heuristic method which is related to the $\chi^2$ statistics discussed above.  Note that David  MacKay~\cite{MacKayBayesNote} explicitly assumes the parameters have an 'intrinsic' arity to them (multinomials with an intrinsic number of bins). This assumption may not be always correct, and in fact, may lead to incorrect assumptions.

Now suppose that the hypotheses are multinomial probability distributions $H_R=\{r_1,\ldots,r_N\}$, with the constraint
that $\sum_i r_i = 1$, and each $r_i$ corresponds to some range (bin) of the data $\mathbf{x}_R$ (and similarly we have $s_i$ for $H_S$), 
then the likelihood ratio can be  written as a sum over the $N$ bins by grouping terms in Equation~\ref{lrat1} into the bins:
\[ \sum_{i\in N} F_i \log\left(\frac{r_i}{s_i}\right)\]
where $F_i$ is the number of data that fall into bin $i$.

The equivalent chi-squared test is to compute the $\chi^2$ statistic for each hypothesis
\[ \chi^2_R=\sum_i \frac{(F_i-r_iN)^2}{r_iN}\;\;\;\;\; \chi^2_S=\sum_i \frac{(F_i-s_iN)^2}{s_iN}\]
and compare them, choosing the one with the smaller $\chi^2$. 

David MacKay argues effectively for the use of the likelihood ratio~\cite{MacKayBayesNote}. We will see in more detail the conditions in which the chi-squared test is not applicable in Section~\ref{sec:gtest}.
}

\mysubsection{Two-Way Likelihood Ratio Test}

If we wish to compare two sets of data, $\mathbf{x}_R$ and $\mathbf{x}_S$, 
and ask whether they are drawn from the same distribution or from two different distributions, then
our first hypothesis is that there are two models $H_R$ and $H_S$ to explain the data, and the second hypothesis is that 
there is a single model $H_{R+S}$ that explains the data. Thus, the question can be formulated as the likelihood ratio
\begin{equation}
L= \log\left(\frac{P(\mathbf{x}_R,\mathbf{x}_S|H_R,H_S)}{P(\mathbf{x}_R,\mathbf{x}_S|H_{R+S})}\right)=\log\left(\frac{P(\mathbf{x}_R|H_R)}{P(\mathbf{x}_R|H_{R+S})}\right)+\log\left(\frac{P(\mathbf{x}_S|H_S)}{P(\mathbf{x}_S|H_{R+S})}\right)
\label{likerat1}
\end{equation}
where we have made the assumption that $\mathbf{x}_R$ is independent of $H_s$ (and vice-versa) if the two distributions are different, and that $\mathbf{x}_R$ is independent of $\mathbf{x}_S$ given $H_{R+S}$ if the two distributions are the same, both of which are true given the i.i.d assumption of data given hypotheses.

\toomuchstats{
The Bayesian formulation of the problem is to parameterise $H_R,H_S$ and $H_{R+S}$ with some unknown parameters, $\theta_R,\theta_S$ and $\theta_{R+S}$, respectively.  The likelihoods in (\ref{likerat1}) are then given by integrating over all possible parameter values
\begin{equation}
L= \log\left(\frac{\int\int P(\theta_R,\theta_S|H_R,H_S)P(\mathbf{x}_R,\mathbf{x}_S|\theta_R,\theta_S,H_R,H_S)d\theta_Rd\theta_S}{\int P(\theta_{R+S}|H_{R+S})P(\mathbf{x}_R,\mathbf{x}_S|\theta_{R+S}H_{R+S})d\theta_{R+s}}\right)
\label{likeratbayes}
\end{equation}
These integrations can sometimes be performed analytically, or using some numerical integration techniques.  However, in this note, we will use the most likely estimate for the parameters, given the data. This simple method is related to the $\chi^2$ statistics discussed above, but will see some limitations of it in Section~\ref{sec:gtest}.
}

We can estimate the parameters of $H_R$ directly from the data, as the most likely estimate using a multinomial with values $r_i=R_{i}/R$, with $R_{i}$ being the number of data points in $\mathbf{x}_R$ that fall into bin $i$, and $R=\sum_i R_{i}$. Similarly for $H_s$ is a multinomial $s_i=S_i/S$, and $S=\sum_i S_{i}$. Finally, we can estimate $H_{R+S}$ in the same way given both datasets, to give a multinomial with values $(R_i+S_i)/(R+S)$.  Using the same transformation (from data to bins) as above, the likelihood ratio becomes
\begin{equation}
L= \sum_{i\in bins} R_i \log\left(\frac{R_i/R}{(R_i+S_i)/(R+S)}\right) + \sum_{i\in bins} S_i \log\left(\frac{S_i/S}{(R_i+S_i)/(R+S)}\right)
\label{gtest}
\end{equation}
which is simply the weighted sum of the Kullback-Leibler divergences of the two datasets from the average distribution
\[L=R\cdot D_{KL}(r_i||p_i)+S\cdot D_{KL}(s_i||p_i)\]
where $p_i=\frac{R_i+S_i}{R+S}$ is the probability of a data point falling in bin $i$ estimated from both sets of data.
It is also a symmetrised relative entropy measure comparing the data to its own distribution (e.g. $R_i$ to $R_i/R$) and to the average
distribution of both sets of data ($(R_i+S_i)/(R+S$). We can see this better by expanding out the logs of fractions as differences of logs
and cancelling terms to obtain.
\[ L=\sum_i \left( R_i\log(\frac{R_i}{R})+ S_i\log(\frac{S_i}{S})  - (R_i+S_i)\log(\frac{R_i+S_i}{R+S})\right)\]
or
\[ L= \left[ R\sum_i r_i\log(r_i) + S \sum_i s_i\log(s_i)  - (R+S)\sum_i p_i\log(p_i)\right]\]
The first term is the (negative) {\em entropy} of the distribution $r_i$ (scaled by the number of datapoints), the second is the negative entropy of $s_i$, and the third is the entropy of the joint distributions.  Denoting $\gamma_r,\gamma_s,\gamma_p$ as the entropy of $r_i$, $s_i$ and $p_i$, respectively, we have
\begin{align}
L &= -\left[R\gamma_r+S\gamma_s-(R+S)\gamma_p\right] \label{likerat-ent} \\
&= -(R+S)\left[\frac{R}{R+S}\gamma_r+\frac{S}{R+S}\gamma_s-\gamma_p\right]\label{likerat-ent2}
\end{align}
where the entropy $\gamma(x)=-x\log(x)$. Equation~\ref{likerat-ent} can be understood by noting that if the two distributions $H_R$ and $H_S$ are the same, then averaging them will make no difference to the entropy of the distributions. If, on the other hand, $H_R$ and $H_S$ are different, then the average of the two will have higher entropy.  Thus, $\gamma_p$ will be larger if the distributions are different, making $L$ also larger (due to the negative sign), which is what we expect from the original definition of the likelihood ratio for the two-way problem as given in (\ref{likerat1}).

More precisely, it is the case that the sum of the entropy of any two probability distributions will be {\em less than} the entropy of their average. To show this, note that the entropy $\gamma(x)=-x\log(x)$ is a {\em concave} function, meaning every point on every {\em chord} lies on or below the function~\cite{BishopBook06}, so that 
\[\alpha\gamma(r)+\beta\gamma(s)  \leq \gamma(\alpha r + \beta s) \]
where $\alpha+\beta=1$, and equality is achieved when $r=s$.  By induction, this is true even for a weighted sum:
\begin{equation}
\alpha \sum_i r_i \log(r_i) + \beta \sum_i s_i log(s_i) \leq \sum_i (\alpha r_i + \beta s_i)\log(\alpha r_i + \beta s_i)
\label{ineq}
\end{equation}
 If we use $\alpha=\frac{R}{R+S}$ and $\beta=\frac{S}{R+S}$, then $p_i=\alpha r_i + \beta s_i$, and Equation~(\ref{ineq})  says that the square bracket in Equation~(\ref{likerat-ent2}) is always negative, so that $L\geq 0$. The extreme cases are 
\begin{enumerate}
\item $r_i$ and $s_i$ are identical, then $L=0$.
\item $r_i=0$ for all $i$ where $s_i>0$, and $s_i=0$ for all $i$ where $r_i>0$.  In this case, either $r_i$ or $s_i$ is zero, and 
\begin{align*}
L&=-(R+S)\left[\alpha \log (\alpha)\sum_i r_i+\beta \log (\beta)\sum_i s_i\right] \\
&= -(R+S)\left[\alpha\log(\alpha)+\beta\log(\beta)\right]
\end{align*}
Since $\alpha+\beta=1$, this function has a maximum of $(R+S)/2$ at $\alpha=0.5$, and a minimum of $0$ at $\alpha=1$ or $0$.  
\end{enumerate}
Thus, we can see that $0\leq L \leq \frac{1}{2}(R+S)$, with the minimum achieved for identical distributions, and the maximum achieved for maximally different distributions.

\toomuchstats{
\mysubsection{Two-Way $\chi^2$ test}
If instead, we use the two-way $\chi^2$ test, we compute the expected counts, which is the average distribution of the two datasets. 
Since $\frac{R_i+S_i}{R+S}$ is the average distribution given both sets of data, we have the expected counts in bin $i$ for the two datasets as
\begin{equation} 
E_R(i) = R\frac{R_i+S_i}{R+S}\;\;\;\;\;\;E_S(i) = S\frac{R_i+S_i}{R+S}
\label{expcounts}
\end{equation}
In many treatments of this problem, particularly in the biological sciences, the $i\in \{1,\ldots,N\}$ are referred to as the {\em rows} and the datasets $\{R,S\}$ are referred to as the {\em columns} in a {\em contingency table}. Typically, the rows are a set of features of the data, and the columns are two different datasets, usually obtained in two different conditions. 

To answer the question of whether the two datasets are drawn from the same hypothesis or not, we formulate the {\em null} hypothesis, which states that they are, and then figure out the expected counts as above. The chi-squared statistic for the two sets of data is
\[ \chi^2 = \sum_{J\in\{R,S\}} \sum_{i\in N} \frac{(J_i-E_J(i))^2}{E_J(i)} = \sum_{i\in N} \frac{(R_i-E_R(i))^2}{E_R(i)}+ \sum_{i\in N} \frac{(S_i-E_S(i))^2}{E_S(i)}\]
putting in the definitions of the expected counts from (\ref{expcounts}) above, and doing some algebra, we get
\[  \chi^2 =\sum_i \frac{\left(\sqrt{S/R}R_i-\sqrt{R/S}S_i^2\right)^2}{R_i+S_i}\]
exactly equation (14.3.3) in~\cite{NumRec}.

This value of $\chi^2$, if large, tells us that the null hypothesis can be rejected, and thus that the distributions are likely to be different. To know what ``large'' means, we can use a chi-squared probability test, that gives us the probability that the sum of the squares of $\nu$ random {\em normal} variables of unit variance and zero mean will be greater than $\chi^2$~\cite{NumRec}. Another way to say this is the probability that a particular value of $\chi^2$ would have occurred by chance if the null hypothesis was correct.  The chi-squared probability test is therefore simply the integral of the probability density of the $\chi^2$ distribution:
\[ P(\chi^2|\nu)=Q(\frac{\nu}{2},\frac{\chi^2}{2})=\frac{\Gamma(\frac{\nu}{2},\frac{\chi^2}{2})}{\Gamma(\frac{\nu}{2})}\]
The number of degrees of freedom in the hypotheses is $\nu$. If the two datasets are drawn without regard for each other (no constraints on the number of datapoints drawn), then the number of degrees of freedom, $\nu$, is the number of bins in which one of the datasets has at least one count.  Typically, if $P(\chi^2|\nu) < 0.05$ (the ``p-value''), the chi-squared test is deemed {\em significant}, and the null hypothesis can be safely rejected.   A simple test that can be used is to reject the null hypothesis if $\chi^2 > \nu$~\cite{NumRec}(p661).

\mysubsection{One- and Two-Way G-test}
\label{sec:gtest}
Interestingly, the likelihood ratio can be more formally related to the $\chi^2$ test, by considering the G-test, defined as~\cite{Sokal94}
\[ G=2\sum_i O_i \log(O_i/E_i)\]
where $O_i$ is the observed counts and $E_i$ is the expected counts. Note that this is simply the Kullback-Leibler divergence between observed and expected counts, multiplied by a factor of two. When summed over all data points in our two-column example, this is
\begin{equation} 
G=2\sum_i R_i \log(\frac{R_i}{E_R(i)}) + 2\sum_i S_i\log(\frac{S_i}{E_S(i)})
\label{eqn:gtest}
\end{equation}
putting in the expressions for the expected counts from above (\ref{expcounts}), we obtain exactly $G=2L$, given by Equation~(\ref{gtest}) above.
 In general, with smaller amounts of data, the chi-squared test will sometimes give incorrect answers, whereas the G-test will not, and so is the recommended test~\cite{MacKayBayesNote,Sokal94}.  To see in more detail why this is so, we can write $O_i=E_i+\delta_i$, with $\sum_i \delta_i=0$ so that the total number of counts stays the same.  The G-test is then
\[ G=2\sum_i (E_i+\delta_i)\log(1+\frac{\delta_i}{E_i}).\] 
If we Taylor expand this around $\frac{\delta_i}{E_i}=0$ (the point at which $O_i$ and $E_i$ agree), and using $\log(1+x)\approx x-\frac{x^2}{2}+O(x^3)$,  we get
\begin{align*}
 G&\approx 2\sum_i (E_i+\delta_i)(\frac{\delta_i}{E_i}-\frac{1}{2}\frac{\delta_i^2}{E_i^2}+O(\delta_i^3)\\
&=2\sum_i \delta_i+\frac{1}{2}\frac{\delta_i^2}{E_i}+O(\delta_i^3)\\
&\approx \sum_i \frac{(O_i-E_i)^2}{E_i}
\end{align*}
and so, we see that $G\approx \chi^2$ when $O_i$ is close to $E_i$. However, the more $O_i$ and $E_i$ are different, the less
well this approximation will work, and $\chi^2$ will tend to compute erroneous answers. The effects of a single outlier in a small
sample set will be more pronounced, which explains why the $\chi^2$ often fails in situations with little data. This is the same
reason why a linear regression can fail with little data, due to the strong effects of outliers.

Since the $\chi^2$ value is just an approximation to the G-value, the G-value can also be used in the chi-squared probability test. This method is recommended by most texts on statistics for the biological sciences. However, it is unclear why one would want to do this, and what the validity is since the chi-squared test is based on the pdf of $\chi^2$.  The G-test directly gives (twice) the log likelihood of the ratio of one hypothesis vs. the other, and so a significance can be attributed directly.  However, recall that these tests are both based on models or hypotheses whose parameters are derived from the data itself. Instead of computing Equation~(\ref{likeratbayes}) directly, as we should do, we are taking the most likely estimate of the parameters $\theta_R,\theta_S$ and $\theta_{R+S}$ (those derived directly from the data), and collapsing the integrals to these point estimates.   One implication of this is that the G-values will depend on the complexity of our models (e.g. the number of bins in our multinomials/histograms). This is simply the model overfitting the data: the models derived from each data set $R$ and $S$ will, with enough complexity, perfectly fit the data.  Therefore, to interpret the G-value from Equation~(\ref{eqn:gtest}), we must take the complexity of the model into account.  To evaluate significance, the value of the likelihood ratio (G/2) should be compared to the number of degrees of freedom, $\nu$. If $G > 2\nu$, then the null hypothesis can be safely rejected. This corresponds roughly to a $p<0.05$. 
}


\mysubsection{Likelihood ratio tests for dynamic models}
In the previous sections, we assumed the data were i.i.d distributed, and that the models (hypotheses) were simple multinomials. It is also possible that the data are sequentially dependent, such as when they come from a dynamic model.  For example, if the data arise from a hidden Markov model, then the same considerations apply as above. For any type of model $H_J,J\in\{R,S,R+S\}$ trained on the data in $J$, we can compute each of $P(\mathbf{x}_R|H_R)$, $P(\mathbf{x}_S|H_S)$, $P(\mathbf{x}_R|H_{R+S})$ and $P(\mathbf{x}_S|H_{R+S})$, and then use  Equation~(\ref{likerat1})  to compute the likelihood ratio, and use a chi-squared probability test as usual.  If the $H$ are hidden Markov models, then the likelihoods will be computed using the standard forward equations~\cite{Dempster77}.

\fabulous{
\subsection*{Acknowledgements}
Thanks to Chris Williams for explaining the factor of $2$ in $G$ and its relationship to $\chi^2$,  to Stephen McKenna for pointing to the Bayesian solution for the problem of integrating over all parameters, which resolves the issue of why a significance test is necessary, and to Olivia Stevenson for pointing out the possibility for emotional creativity.
}
\bibliographystyle{plain}
\bibliography{../../refs/refs}

\begin{thebibliography}{1}

\bibitem{BishopBook06}
Christopher~M. Bishop.
\newblock {\em Pattern Recognition and Machine Learning}.
\newblock Springer, 2006.

\bibitem{Dempster77}
A.P. Dempster, N.M. Laird, and D.B. Rubin.
\newblock Maximum likelihood from incomplete data using the {EM} algorithm.
\newblock {\em Journal of the Royal Statistical Society}, 39(B):1--38, 1977.

\bibitem{MacKayBayesNote}
David~J.C. MacKay.
\newblock Bayes or chi-squared? or does it not matter?, 2005.

\bibitem{NumRec}
William~H. Press, Saul~A. Teukolsky, William~T. Vetterling, and Brian~P.
  Flannery.
\newblock {\em Numerical Recipies in C}.
\newblock Cambridge University Press, 2 edition, 1992.

\bibitem{Sokal94}
Robert~R. Sokal and F.~James Rohlf.
\newblock {\em Biometry: The Principles and Practices of Statistics in
  Biological Research}.
\newblock W.H. Freeman, 3 edition, 1994.

\end{thebibliography}

\end{document}